\pdfoutput=1
\PassOptionsToPackage{unicode}{hyperref}
\PassOptionsToPackage{hyphens}{url}
\PassOptionsToPackage{dvipsnames,svgnames,x11names}{xcolor}
\documentclass[
  12pt]{article}

\usepackage{amsmath,amssymb}
\usepackage[T1]{fontenc}
\usepackage[utf8]{inputenc}
\usepackage{textcomp} 
\usepackage{lmodern}
\IfFileExists{upquote.sty}{\usepackage{upquote}}{}
\IfFileExists{microtype.sty}{
  \usepackage[]{microtype}
  \UseMicrotypeSet[protrusion]{basicmath} 
}{}
\makeatletter
\@ifundefined{KOMAClassName}{
  \IfFileExists{parskip.sty}{%
    \usepackage{parskip}
  }{
    \setlength{\parindent}{0pt}
    \setlength{\parskip}{6pt plus 2pt minus 1pt}}
}{
  \KOMAoptions{parskip=half}}
\makeatother
\usepackage{xcolor}
\makeatletter
\ifx\paragraph\undefined\else
  \let\oldparagraph\paragraph
  \renewcommand{\paragraph}{
    \@ifstar
      \xxxParagraphStar
      \xxxParagraphNoStar
  }
  \newcommand{\xxxParagraphStar}[1]{\oldparagraph*{#1}\mbox{}}
  \newcommand{\xxxParagraphNoStar}[1]{\oldparagraph{#1}\mbox{}}
\fi
\ifx\subparagraph\undefined\else
  \let\oldsubparagraph\subparagraph
  \renewcommand{\subparagraph}{
    \@ifstar
      \xxxSubParagraphStar
      \xxxSubParagraphNoStar
  }
  \newcommand{\xxxSubParagraphStar}[1]{\oldsubparagraph*{#1}\mbox{}}
  \newcommand{\xxxSubParagraphNoStar}[1]{\oldsubparagraph{#1}\mbox{}}
\fi
\makeatother

\usepackage{longtable,booktabs,array}
\usepackage{calc} 
\usepackage{etoolbox}
\makeatletter
\patchcmd\longtable{\par}{\if@noskipsec\mbox{}\fi\par}{}{}
\makeatother
\IfFileExists{footnotehyper.sty}{\usepackage{footnotehyper}}{\usepackage{footnote}}
\makesavenoteenv{longtable}
\usepackage{graphicx}
\makeatletter
\def\maxwidth{\ifdim\Gin@nat@width>\linewidth\linewidth\else\Gin@nat@width\fi}
\def\maxheight{\ifdim\Gin@nat@height>\textheight\textheight\else\Gin@nat@height\fi}
\makeatother
\setkeys{Gin}{width=\maxwidth,height=\maxheight,keepaspectratio}
\makeatletter
\def\fps@figure{htbp}
\makeatother

\makeatletter
\@ifpackageloaded{caption}{}{\usepackage{caption}}
\AtBeginDocument{%
\ifdefined\contentsname
  \renewcommand*\contentsname{Table of contents}
\else
  \newcommand\contentsname{Table of contents}
\fi
\ifdefined\listfigurename
  \renewcommand*\listfigurename{List of Figures}
\else
  \newcommand\listfigurename{List of Figures}
\fi
\ifdefined\listtablename
  \renewcommand*\listtablename{List of Tables}
\else
  \newcommand\listtablename{List of Tables}
\fi
\ifdefined\figurename
  \renewcommand*\figurename{Figure}
\else
  \newcommand\figurename{Figure}
\fi
\ifdefined\tablename
  \renewcommand*\tablename{Table}
\else
  \newcommand\tablename{Table}
\fi
}
\@ifpackageloaded{float}{}{\usepackage{float}}
\floatstyle{ruled}
\@ifundefined{c@chapter}{\newfloat{codelisting}{h}{lop}}{\newfloat{codelisting}{h}{lop}[chapter]}
\floatname{codelisting}{Listing}

\makeatother
\makeatletter
\@ifpackageloaded{caption}{}{\usepackage{caption}}
\@ifpackageloaded{subcaption}{}{\usepackage{subcaption}}
\makeatother

\usepackage[]{natbib}
\usepackage{bookmark}

\IfFileExists{xurl.sty}{\usepackage{xurl}}{} 
\hypersetup{
  pdftitle={Design-based causal inference for ROC curves},
  pdfauthor={Bufan Li; Haoyang Yu},
  pdfkeywords={3 to 6 keywords, that do not appear in the title},
  colorlinks=true,
  linkcolor={blue},
  filecolor={Maroon},
  citecolor={Blue},
  urlcolor={Blue},
  pdfcreator={LaTeX via pandoc}}

\usepackage{amsmath}
\usepackage{graphicx}
\usepackage{enumerate}
\usepackage{natbib}
\usepackage{silence}
\usepackage{bm}
\usepackage{bbm}
\usepackage{url} 
\usepackage{amsthm}
\usepackage{amsfonts}
\usepackage{amssymb}
\usepackage{multirow}
\usepackage{booktabs}
\usepackage[justification=centering]{caption}
\usepackage{threeparttable}
\usepackage{graphicx}
\usepackage{float}
\usepackage[section]{placeins}
\usepackage{comment}
\usepackage{subcaption}
\usepackage{cleveref}
\usepackage{xr}
\usepackage{threeparttable}
\usepackage{xcolor}
\usepackage{colortbl}
\usepackage{titling}
\usepackage{algorithmic}
\usepackage{algorithm}

\usepackage{enumitem}
\usepackage{multibib}
\usepackage{xr}
\makeatletter
\def\@extra@binfo{app}
\newcites{app}{References}
\def\@extra@binfo{}
\makeatother
\bibliographystyleapp{agsm}

\theoremstyle{definition}
\newtheorem{assumption}{Assumption}
\Crefname{assumption}{Assumption}{Assumptions}
\crefrangelabelformat{assumption}{#3#1#4--#5#2#6}
\theoremstyle{plain}
\newtheorem{theorem}{Theorem}
\newtheorem{proposition}{Proposition}

\newtheorem{corollary}{Corollary}
\theoremstyle{remark}

\newcommand{\E}{\mathbb E}
\renewcommand{\Pr}{\mathbb P}

\newcommand{\QTE}{\operatorname{QTE}}
\newcommand{\DTE}{\operatorname{DTE}}
\newcommand{\ind}{\mathbb I}
\newcommand{\AUC}{\operatorname{AUC}}

\newcommand{\anon}{1}

\begin{document}

\def\spacingset#1{\renewcommand{\baselinestretch}%
{#1}\small\normalsize} \spacingset{1}


\if1\anon
{
  \title{\bf Design-based causal inference for ROC curves}
  \author{Bufan Li and Haoyang Yu\thanks{Bufan Li (\texttt{lbf23@mails.tsinghua.edu.cn}) and Haoyang Yu (\texttt{yhy22@mails.tsinghua.edu.cn}) are listed in alphabetical order and contributed equally. We used ChatGPT and Gemini to assist with proofreading and editing during the preparation of this manuscript. All errors are our own.}\hspace{.2cm}\\
  Department of Statistics and Data Science, Tsinghua University}
  \maketitle
} \fi

\if0\anon
{
  \bigskip
  \bigskip
  \bigskip
  \begin{center}
    {\LARGE\bf Design-based causal inference for ROC curves}
\end{center}
  \medskip
} \fi

\bigskip
\begin{abstract}
In randomized experiments, causal inference traditionally focuses on the average treatment effect, yet comparing the treatment and control potential outcome distributions is often of interest.
Under complete randomization, we adopt a design-based framework without imposing an outcome model to study the receiver operating characteristic (ROC) curve, which measures the proportion of treated outcomes exceeding each upper quantile of the control distribution.
We propose an estimator that depends only on pooled outcome ranks, establish its asymptotic normality, and construct conservative variance estimators and confidence intervals.
We also develop covariate adjustment directly for the ROC estimator, preserving the target estimand while achieving an asymptotic variance no larger than that of the unadjusted estimator.
We further construct simultaneous confidence bands over an interior range of quantile levels and extend pointwise inference to the extreme upper tail of the control distribution, with both extensions accommodating covariate adjustment.
Simulation studies and an application to a large-scale online learning experiment illustrate the proposed methods.
\end{abstract}

\noindent%
\vfill

\newpage
\spacingset{1.8} 

\section{Introduction}
\label{sec:intro}

Randomized experiments are widely used to evaluate causal effects, often summarized by the average treatment effect (ATE) \citep{Neyman1923,liding2017,ding2024first}.
In many applications, however, interest extends beyond mean effects to differences across the outcome distribution.
For example, when evaluating an educational intervention, investigators may wish to know the proportions of treated students who outperform the median, upper quartile, or top decile of the control group.
Such comparisons are particularly relevant when outcomes are available only as rankings rather than numerical measurements, making conventional outcome models inapplicable.
Although baseline covariates have been widely used to improve the efficiency of ATE estimation \citep{lin2013,liding2017,zhaodingli2024}, extending covariate adjustment to rank-based inference without imposing outcome models remains challenging.
These considerations motivate two questions:
(i) Can we develop a design-based framework for valid causal inference across the entire outcome distribution using only rank information?
(ii) Can we leverage baseline covariates to improve the efficiency of such inference?

In this paper, we address both questions through the following contributions.
First, we define the finite-population receiver operating characteristic (ROC) curve and develop a rank-based estimator under complete randomization.
Adopting a design-based perspective \citep{imbens2015causal,ding2024first}, we treat potential outcomes as fixed and treatment assignment as the sole source of randomness, without imposing outcome models.
For each $\alpha\in(0,1)$, the causal ROC curve measures the proportion of treatment potential outcomes that exceed the $(1-\alpha)$-quantile of the control potential outcome distribution.
Deviations from the diagonal $R(\alpha)=\alpha$ reveal distributional differences between treatment and control potential outcomes.
Unlike classical ROC analysis, which compares test scores from two populations \citep{swets1988roc,hsiehturnbull1996roc,Pepe2003,Fawcett2006}, the causal ROC curve compares the two potential outcome distributions within the same finite population, where each unit has both treatment and control potential outcomes but only one is observed \citep{Neyman1923,rubin1974}.
We construct a plug-in estimator using only pooled outcome ranks, accommodating both continuous and discrete outcomes without requiring numerical outcome measurements.

Second, we develop a covariate adjustment method to improve the precision of rank-based ROC estimation.
Existing covariate adjustment methods for the ATE \citep{lin2013,liding2017} typically rely on numerical outcomes, while model-based ROC regression methods \citep{BrumbackPepeAlonzo2006} impose assumptions on outcome distributions.
To overcome these limitations, we apply linear covariate adjustment directly to the rank-based ROC estimator.
By exploiting covariate balance under randomization, our method preserves the marginal finite-population ROC estimand without imposing outcome models and achieves an asymptotic variance no larger than that of the unadjusted estimator.

Third, we develop a unified framework for pointwise inference, simultaneous confidence bands, and tail inference, both with and without covariate adjustment.
These three settings address distinct inferential goals: evaluating treatment effects at a prespecified threshold (e.g., the top $10\%$ of controls) calls for \emph{pointwise inference}, comparing distributions across a range of thresholds requires \emph{simultaneous inference}, and examining extreme outcomes (e.g., the top $0.1\%$ of controls) motivates \emph{tail inference}, where the tail probability $\alpha_n\to0$ as the population size $n$ increases.
For pointwise inference, we establish asymptotic normality, construct Neyman-type conservative variance estimators and confidence intervals \citep{Neyman1923}, and propose two estimators of the unknown local ROC slope, one based on density ratios and the other using only pooled outcome ranks.
For simultaneous inference, we establish a functional central limit theorem and construct conservative simultaneous confidence bands over a fixed interior interval using Gaussian simulation.
For tail inference, where increasingly sparse observations make conventional local slope estimators unreliable, we propose a new tail slope estimator, establish asymptotic normality, and construct conservative confidence intervals.

Finally, we demonstrate the finite-sample performance of our methods and the practical utility of our framework through simulation studies and an application to a large-scale randomized educational experiment \citep{asanov2023}.

Our work relates to several lines of research on distributional and rank-based causal inference. Quantile treatment effects (QTEs) compare outcome quantiles at common percentiles \citep{ZhangZheng2020,hanwang2026}, while distributional treatment effects (DTEs) compare cumulative probabilities at fixed outcome thresholds \citep{ByambadalaiOkaYasui2024,OkaEtAl2026}. By defining thresholds through control quantiles, the ROC curve instead characterizes relative performance across the control distribution and accommodates rank-only outcomes. Unlike the scalar area under the ROC curve (AUC), which summarizes the probability that a treated outcome exceeds a control outcome \citep{chen2026design}, the full ROC curve reveals distributional differences across performance tiers that a single summary cannot capture \citep{Hand2009}.

Previous ROC research has studied pointwise regression under superpopulation models \citep{BrumbackPepeAlonzo2006} and simultaneous confidence bands and empirical-process asymptotics for independent two-sample settings \citep{MaHall1993,HorvathEtAl2008,KoopmeinersFeng2011}. In contrast, we develop a model-free, design-based framework under finite-population randomization that integrates pointwise, simultaneous, and shrinking-tail inference with covariate adjustment. Our tail analysis connects to the literature on high-quantile estimation and extremal quantile treatment effects \citep{DekkersdeHaan1989,Chernozhukov2005,WangLiHe2012,Zhang2018}. While these studies primarily examine numerical outcome tails under superpopulation frameworks, we develop design-based inference for relative tail performance on the control-percentile scale, requiring only pooled ranks.


\textbf{Notation.} Finite-population quantities carry a subscript $n$, while their limiting counterparts omit it.
Write $\ind(\cdot)$ for the indicator function, $\|\cdot\|$ for the Euclidean norm, and $a_n\asymp b_n$ when $a_n/b_n$ is bounded away from zero and infinity.
For a nondecreasing function $F$, define $F^{-1}(v)=\inf\{y\in\mathbb R:F(y)\ge v\}$, and for a nonincreasing function $G$, define $G^{-1}(\alpha)=\inf\{y\in\mathbb R:G(y)\le\alpha\}$.
For arrays $\{\bm g_i\}_{i=1}^n$ and $\{\bm h_i\}_{i=1}^n$, define $\bar{\bm g}_n=n^{-1}\sum_{i=1}^n\bm g_i$ and $\bm S_{\bm g\bm h,n}=(n-1)^{-1}\sum_{i=1}^n(\bm g_i-\bar{\bm g}_n)(\bm h_i-\bar{\bm h}_n)^\top$, where $\bar{\bm h}_n$ is defined analogously.
When both arrays are scalar-valued, we write $\bar g_n$ and $S_{gh,n}$; when only $h_i$ is scalar, we write $\bm S_{\bm g h,n}$ for the corresponding covariance vector.

\section{Framework}
\label{sec:framework}

\subsection{ROC curve}
\label{sec:roc}

Consider a completely randomized experiment on a finite population of $n$ units. For each unit $i\in\{1,\ldots,n\}$, let $Y_i(1)$ and $Y_i(0)$ denote its potential outcomes under treatment and control, $Z_i\in\{0,1\}$ its treatment assignment, and $\bm X_i\in\mathbb R^d$ its pre-treatment covariates. Exactly $n_1$ units are selected uniformly at random to receive treatment ($Z_i=1$), while the remaining $n_0=n-n_1$ serve as controls ($Z_i=0$). Let $\bm Z=(Z_1,\ldots,Z_n)^\top$ denote the random treatment assignment vector and $\bm z=(z_1,\ldots,z_n)^\top$ a possible realization. The randomization distribution is given by
\begin{align*}
\Pr(\bm Z=\bm z)=\binom{n}{n_1}^{-1},
\quad \bm z\in\{0,1\}^n \text{ with } \sum_{i=1}^n z_i=n_1.
\end{align*}
Throughout the paper, we assume that the treatment proportion $p_n=n_1/n$ converges to a constant $p\in(0,1)$ as $n\to\infty$. We adopt a design-based framework that treats the finite population $\{Y_i(1),Y_i(0),\bm X_i\}_{i=1}^n$ as fixed, with all randomness arising from the treatment assignment vector $\bm Z$, without imposing an outcome model.

For each $z\in\{0,1\}$ and $y\in\mathbb R$, define
\begin{align*}
F_{z,n}(y)=\frac1n\sum_{i=1}^n\ind\{Y_i(z)\le y\},\quad
G_{z,n}(y)=1-F_{z,n}(y),\quad
m_{z,n}(y)=\frac1n\sum_{i=1}^n\ind\{Y_i(z)=y\},
\end{align*}
as the finite-population cumulative distribution function, survival function, and probability mass function under treatment $z$, respectively.

To compare the two potential outcome distributions across thresholds, we introduce the receiver operating characteristic (ROC) curve, which characterizes the relationship between the probabilities of treatment and control outcomes exceeding a common threshold as the threshold varies.

Specifically, let $F_0$ and $F_1$ denote two continuous cumulative distribution functions, with corresponding survival functions $G_z(y)=1-F_z(y)$ for $z\in\{0,1\}$. The ROC curve is defined as 
\begin{align*}
  R(\alpha)=G_1\{G_0^{-1}(\alpha)\},\quad \alpha\in(0,1).
\end{align*}

In a finite population, however, the survival functions $G_{0,n}$ and $G_{1,n}$ are necessarily discrete. Although the ROC curve can still be defined as above, $G_{0,n}\{G_{0,n}^{-1}(\alpha)\}\neq\alpha$ in general. Consequently, $\alpha$ no longer represents the control exceedance probability corresponding to the threshold $G_{0,n}^{-1}(\alpha)$, complicating the interpretation of the ROC curve.
To ensure that the control exceedance probability equals the specified level $\alpha$, we allow fractional inclusion of outcomes at the threshold. Specifically, for $z\in\{0,1\}$, define
\begin{align*}
W_i(z;t,\lambda)
=\ind\{Y_i(z)>t\}+\lambda\ind\{Y_i(z)=t\},
\qquad \lambda\in[0,1],
\end{align*}
where $\lambda$ determines the fraction of observations tied at $t$ that are counted as exceeding the threshold.

For each $\alpha\in(0,1)$, let $q_n(\alpha)=G_{0,n}^{-1}(\alpha)$ denote the control upper-tail quantile and choose $\lambda_n(\alpha)\in[0,1]$ such that
\begin{align*}
\frac1n\sum_{i=1}^n W_i\{0;q_n(\alpha),\lambda_n(\alpha)\}=\alpha.
\end{align*}
Solving this equation yields
\begin{align*}
\lambda_n(\alpha)
=\frac{\alpha-G_{0,n}\{q_n(\alpha)\}}{m_{0,n}\{q_n(\alpha)\}}.
\end{align*}
We define the finite-population ROC curve as
\begin{equation}
R_n(\alpha)=\frac1n\sum_{i=1}^n W_i\{1;q_n(\alpha),\lambda_n(\alpha)\}
=G_{1,n}\{q_n(\alpha)\}+\lambda_n(\alpha)m_{1,n}\{q_n(\alpha)\},
\label{eq:roc}
\end{equation}
where $R_n(\alpha)$ represents the weighted proportion of treatment potential outcomes exceeding the control upper-tail quantile $q_n(\alpha)$, with ties weighted by $\lambda_n(\alpha)$. If the two marginal distributions coincide, i.e., $F_{1,n}(t)=F_{0,n}(t)$ for all $t\in\mathbb R$, then $R_n(\alpha)=\alpha$ for all $\alpha\in(0,1)$.

The ROC curve depends only on the relative ordering of the potential outcomes, including ties, and is therefore invariant to any common strictly increasing transformation. Consequently, it can be identified from pooled ranks of the observed treatment and control outcomes, without requiring their numerical values, as established in Proposition~\ref{p:rank} below.

\subsection{Relation to the AUC and average, quantile and distributional treatment effects}
\label{sec:compare}

To clarify the relationship between the causal ROC curve and existing causal estimands, we compare it with four related measures: the average treatment effect (ATE) \citep{Neyman1923,imbens2015causal,liding2017}, the area under the ROC curve (AUC) \citep{HanleyMcNeil1982}, the quantile treatment effect (QTE) \citep{ZhangZheng2020,hanwang2026}, and the distributional treatment effect (DTE) \citep{ByambadalaiOkaYasui2024,OkaEtAl2026}.
These measures are defined as follows:
\begin{align*}
\tau_n&=\frac1n\sum_{i=1}^n\{Y_i(1)-Y_i(0)\}
=\int_{\mathbb R}\{F_{0,n}(y)-F_{1,n}(y)\}\,\mathrm{d}y
=\int_{\mathbb R}\{G_{1,n}(y)-G_{0,n}(y)\}\,\mathrm{d}y,\\
\AUC_n&=\int_0^1 R_n(\alpha)\,\mathrm{d}\alpha=\frac{1}{n^2}\sum_{i=1}^n\sum_{j=1}^n
\left[
\ind\{Y_i(1)>Y_j(0)\}
+\frac12\ind\{Y_i(1)=Y_j(0)\}
\right],
\end{align*}
and, for $v\in(0,1)$ and $y\in\mathbb R$, by
\begin{align*}
\QTE_n(v)&=F_{1,n}^{-1}(v)-F_{0,n}^{-1}(v),\\
\DTE_n(y)&=\{1-F_{1,n}(y)\}-\{1-F_{0,n}(y)\}=G_{1,n}(y)-G_{0,n}(y).
\end{align*} 

These measures address different questions about treatment effects. Both the ATE and the AUC, which is closely related to the Mann--Whitney parameter studied under design-based inference by \citet{chen2026design}, collapse the comparison into a single scalar summary (mean difference vs.\ overall pairwise ordering) and can therefore mask heterogeneous effects across outcome levels \citep{Hand2009}. Among functional measures, the QTE compares outcome levels at a common percentile, whereas both the DTE and the ROC curve compare exceedance probabilities. Specifically, for continuous limiting distributions with $q(\alpha)=G_0^{-1}(\alpha)$, they satisfy the algebraic correspondence
\begin{align*}
R(\alpha)-\alpha=G_1\{q(\alpha)\}-G_0\{q(\alpha)\}=\DTE\{q(\alpha)\}.
\end{align*}
Despite this connection, the two estimands differ fundamentally in their estimation and inference: the DTE evaluates exceedance at a fixed numerical threshold $y$ through a difference in sample proportions, whereas the ROC curve evaluates exceedance at an unknown control quantile $q_n(\alpha)$ that must be estimated from the data.
Unlike the ATE, QTE, and DTE, both the AUC and the ROC curve can be estimated using only pooled outcome ranks, without requiring numerical outcome values.
Table~\ref{tab:estimands} summarizes these distinctions.

\begin{table}[!htbp]
\centering
\caption{Comparison of five treatment-effect measures.}
\label{tab:estimands}
\begin{tabular}{llll}
\toprule
Measure & Comparison & Reference standard & Rank-only outcomes \\
\midrule
ATE & Mean outcomes & None & No \\
AUC & Pairwise ordering & None & Yes \\
QTE & Outcome quantiles & Common percentile $v$ & No \\
DTE & Exceedance probabilities & Fixed outcome level $y$ & No \\
ROC & Exceedance probabilities & Control-relative level $\alpha$ & Yes \\
\bottomrule
\end{tabular}
\end{table}

\begin{figure}[!t]
\centering
\includegraphics[width=0.92\textwidth]{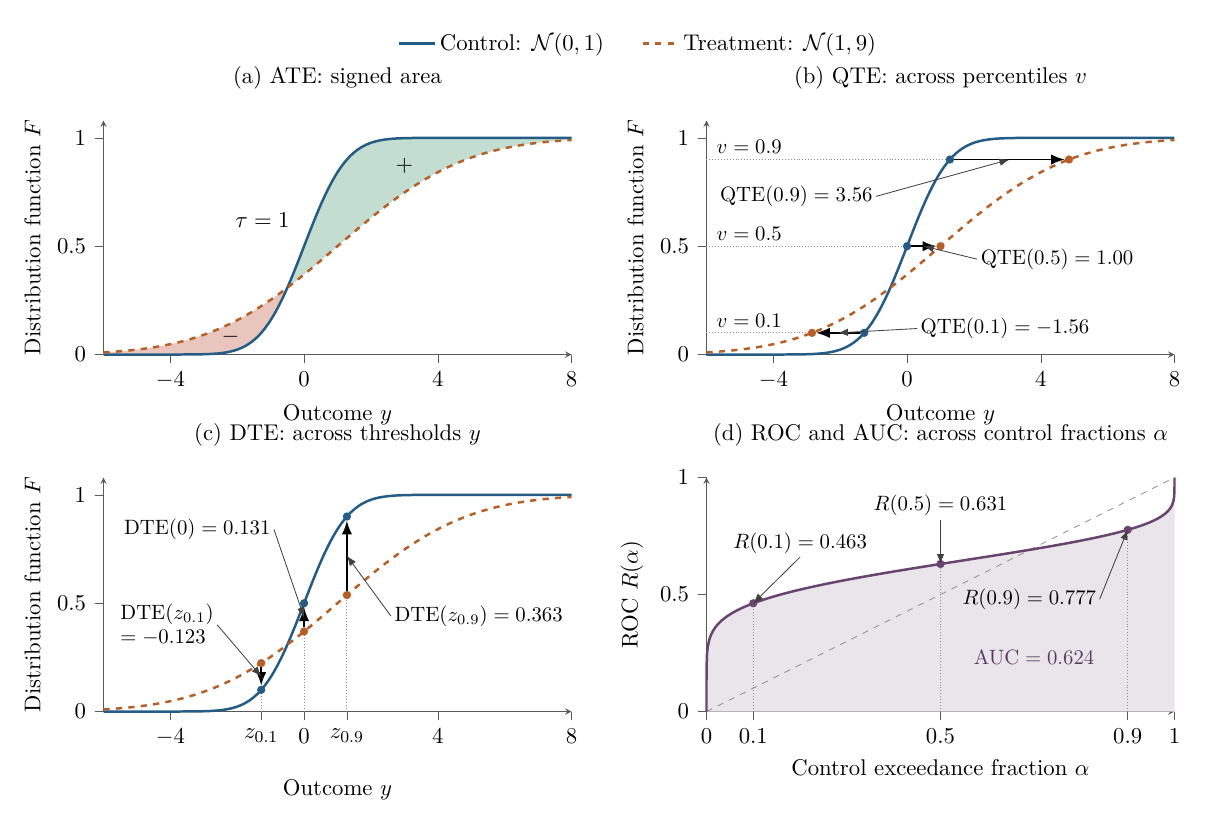}
\caption{Illustration of five treatment-effect measures.}
\label{fig:estimand-geometry}
\end{figure}

To illustrate these five measures, suppose that the control and treatment outcomes follow $\mathcal N(0,1)$ and $\mathcal N(1,9)$, respectively.
Let $\Phi$ denote the standard normal cumulative distribution function and $z_v=\Phi^{-1}(v)$.
Figure~\ref{fig:estimand-geometry} illustrates these measures at three levels of the outcome distributions.
Panel~(a) shows the signed area corresponding to $\tau=1$. Panels~(b) and~(c) reveal this heterogeneity through $\QTE(0.1)=-1.56$, $\QTE(0.5)=1.00$, $\QTE(0.9)=3.56$, and $\DTE(z_{0.1})=-0.123$, $\DTE(0)=0.131$, $\DTE(z_{0.9})=0.363$. Panel~(d) gives the corresponding ROC values $R(0.9)=0.777$, $R(0.5)=0.631$, and $R(0.1)=0.463$, with the shaded area giving $\AUC=0.624$. While the ATE and AUC provide scalar summaries, the QTE, DTE, and ROC curve characterize distributional differences across different levels of the outcome distributions.

\section{Pointwise inference}
\label{sec:inference}

\subsection{Estimation}
\label{sec:estimators}

Throughout this section, we fix $\alpha\in(0,1)$ and develop pointwise estimation and inference for the finite-population ROC curve. For each $z\in\{0,1\}$, recall that
\begin{align*}
G_{z,n}(y)=\frac1n\sum_{i=1}^n\ind\{Y_i(z)>y\},\qquad
m_{z,n}(y)=\frac1n\sum_{i=1}^n\ind\{Y_i(z)=y\}.
\end{align*}
Let $Y_i=Z_iY_i(1)+(1-Z_i)Y_i(0)$ denote the observed outcome of unit $i$. Under complete randomization, unbiased estimators of $G_{z,n}(y)$ and $m_{z,n}(y)$ can be given by
\begin{align*}
\widehat G_{z,n}(y)=\frac1{n_z}\sum_{i:Z_i=z}\ind\{Y_i>y\},\qquad
\widehat m_{z,n}(y)=\frac1{n_z}\sum_{i:Z_i=z}\ind\{Y_i=y\}.
\end{align*}
Replacing $G_{0,n}$ and $m_{0,n}$ with their sample counterparts yields the plug-in estimators
\begin{align*}
\widehat q_n(\alpha)=\widehat G_{0,n}^{-1}(\alpha),
\qquad
\widehat\lambda_n(\alpha)
=\frac{\alpha-\widehat G_{0,n}\{\widehat q_n(\alpha)\}}{\widehat m_{0,n}\{\widehat q_n(\alpha)\}}\in[0,1].
\end{align*}
Substituting these into \eqref{eq:roc} gives
\begin{equation}
\widehat R_n(\alpha)=\widehat G_{1,n}\{\widehat q_n(\alpha)\}+\widehat\lambda_n(\alpha)\widehat m_{1,n}\{\widehat q_n(\alpha)\}.
\label{eq:roc-hat}
\end{equation}
The estimator in \eqref{eq:roc-hat} is computed from numerical outcome values. Proposition~\ref{p:rank} shows that the ROC estimator can also be computed from pooled ranks alone.

\begin{proposition}
\label{p:rank}
Let $\mathrm{rank}_{i,n}$ denote the rank of $Y_i$ among $\{Y_1,\ldots,Y_n\}$, with ties assigned equal ranks. Let $\widehat R_n^{\mathrm{rank}}(\alpha)$ denote the estimator in \eqref{eq:roc-hat} computed using $\mathrm{rank}_{i,n}$ in place of $Y_i$. Then, for every $\alpha\in(0,1)$, $\widehat R_n^{\mathrm{rank}}(\alpha)=\widehat R_n(\alpha)$.
\end{proposition}

Unless otherwise stated, we formulate the subsequent results in terms of the original outcomes and discuss rank-based implementations when necessary.

\subsection{Asymptotic normality}
\label{sec:normal}

\begin{assumption}
\label{a:margins}
There exist distribution functions $F_0$ and $F_1$ such that
\begin{align*}
\sup_{y\in\mathbb R}|F_{z,n}(y)-F_z(y)|\to0,\qquad z=0,1.
\end{align*}
Write $G_z=1-F_z$ for $z=0,1$.
\end{assumption}

\begin{assumption}
\label{a:local}
For a fixed $\alpha\in(0,1)$, there exist $q(\alpha)\in\mathbb R$ and positive constants $f_0\{q(\alpha)\}$ and $f_1\{q(\alpha)\}$ such that $q_n(\alpha)\to q(\alpha)$, $G_0\{q(\alpha)\}=\alpha$, and, for every fixed $M>0$,
\begin{align*}
\sup_{|t|\le M}\Bigl|\sqrt n\bigl[G_{z,n}\{q_n(\alpha)+t/\sqrt n\}-G_{z,n}\{q_n(\alpha)\}\bigr]+tf_z\{q(\alpha)\}\Bigr|\to0,
\qquad z=0,1.
\end{align*}
\end{assumption}

Assumption~\ref{a:margins} requires uniform convergence of the finite-population marginal distributions, as also considered by \citet{du2026sharp}.
Assumption~\ref{a:local} requires local linear approximations of the finite-population marginal distributions around the control quantile at the $n^{-1/2}$ scale, analogous to the condition used by \citet{hanwang2026} for design-based inference on quantile treatment effects.

Under Assumptions~\ref{a:margins} and~\ref{a:local}, the estimation error of $\widehat R_n(\alpha)$ admits the following asymptotic linear expansion, proved in Section~\ref{sec:app-linear} of the Supplementary Material:
\begin{align*}
\begin{aligned}
\widehat R_n(\alpha)-R_n(\alpha)
&=\bigl[\widehat G_{1,n}\{q_n(\alpha)\}-G_{1,n}\{q_n(\alpha)\}\bigr]\\
&\quad-r(\alpha)\bigl[\widehat G_{0,n}\{q_n(\alpha)\}-G_{0,n}\{q_n(\alpha)\}\bigr]
+o_P(n^{-1/2}),
\end{aligned}
\end{align*}
where $r(\alpha)=f_1\{q(\alpha)\}/f_0\{q(\alpha)\}$ is the local ROC slope, which extends the derivative of the ROC curve to the finite-population setting.
The first term captures the sampling fluctuation of the treatment survival function evaluated at the population threshold $q_n(\alpha)$.
The second term accounts for the uncertainty in estimating the control quantile $q_n(\alpha)$, with the local ROC slope $r(\alpha)$ determining how this uncertainty propagates into the ROC estimator.
Define the centered treatment- and control-arm linearization terms by
\begin{equation}
\begin{aligned}
\psi_{i,1,n}(\alpha)&=\ind\{Y_i(1)>q_n(\alpha)\}-G_{1,n}\{q_n(\alpha)\},\\
\psi_{i,0,n}(\alpha)&=r(\alpha)\bigl[\ind\{Y_i(0)>q_n(\alpha)\}-G_{0,n}\{q_n(\alpha)\}\bigr].
\end{aligned}
\label{eq:linearization-terms}
\end{equation}
By construction, $\sum_{i=1}^n\psi_{i,z,n}(\alpha)=0$ for $z=0,1$. 
The preceding expansion can then be expressed as
\begin{equation}
\widehat R_n(\alpha)-R_n(\alpha)=L_n(\alpha)+o_P(n^{-1/2}),
\qquad
L_n(\alpha)=\frac1{n_1}\sum_{i:Z_i=1}\psi_{i,1,n}(\alpha)-\frac1{n_0}\sum_{i:Z_i=0}\psi_{i,0,n}(\alpha).
\label{eq:linear}
\end{equation}
Let $\delta_{i,n}(\alpha)=\psi_{i,1,n}(\alpha)-\psi_{i,0,n}(\alpha)$ and $\bar\psi_{z,n}(\alpha)=n_z^{-1}\sum_{i:Z_i=z}\psi_{i,z,n}(\alpha)$ for $z=0,1$. Since $\psi_{i,1,n}(\alpha)$, $\psi_{i,0,n}(\alpha)$, and $\delta_{i,n}(\alpha)$ all have population mean zero, let $S_{\psi_z\psi_z,n}(\alpha)=(n-1)^{-1}\sum_{i=1}^n\psi_{i,z,n}(\alpha)^2$ ($z=0,1$), $S_{\psi_1\psi_0,n}(\alpha)=(n-1)^{-1}\sum_{i=1}^n\psi_{i,1,n}(\alpha)\psi_{i,0,n}(\alpha)$, and $S_{\delta\delta,n}(\alpha)=(n-1)^{-1}\sum_{i=1}^n\delta_{i,n}(\alpha)^2$ denote their finite-population variances and covariance.
The randomization variance of $\sqrt n L_n(\alpha)$ is then
\begin{align*}
\sigma_{R,n}^2(\alpha)
=\frac{S_{\psi_1\psi_1,n}(\alpha)}{p_n}
+\frac{S_{\psi_0\psi_0,n}(\alpha)}{1-p_n}
-S_{\delta\delta,n}(\alpha).
\end{align*}

\begin{assumption}
\label{a:limits}
The finite-population variances $S_{\psi_1\psi_1,n}(\alpha)$ and $S_{\psi_0\psi_0,n}(\alpha)$ and covariance $S_{\psi_1\psi_0,n}(\alpha)$ converge to finite limits, with $\sigma_{R,n}^2(\alpha)\to\sigma_R^2(\alpha)>0$.
\end{assumption}

Assumption~\ref{a:limits} imposes standard convergence and nondegeneracy conditions on finite-population variances and covariances, as in \citet{liding2017}, but on $\psi_{i,1,n}(\alpha)$ and $\psi_{i,0,n}(\alpha)$ rather than the original potential outcomes $Y_i(1)$ and $Y_i(0)$.
The following \Cref{t:clt} establishes the asymptotic normality of the ROC estimator.

\begin{theorem}
\label{t:clt}
Under Assumptions~\ref{a:margins}--\ref{a:limits}, $\sqrt n\{\widehat R_n(\alpha)-R_n(\alpha)\}
\xrightarrow{d}\mathcal N\{0,\sigma_R^2(\alpha)\}$.
\end{theorem}

\subsection{Variance estimation and confidence intervals}
\label{sec:interval}

Based on Theorem~\ref{t:clt}, constructing confidence intervals for $R_n(\alpha)$ requires estimating $\sigma_{R,n}^2(\alpha)$.
However, the term $S_{\delta\delta,n}(\alpha)$ is not identifiable because the two potential outcomes cannot be jointly observed for any unit. 
Following the classical Neyman approach \citep{Neyman1923,liding2017}, we omit this nonnegative term and estimate the resulting conservative variance bound using within-arm sample variances.

Specifically, given an estimator $\widehat r_n(\alpha)$ of $r(\alpha)$, define
\begin{align*}
\widehat\psi_{i,1,n}(\alpha)
&=\ind\{Y_i>\widehat q_n(\alpha)\}
-\widehat G_{1,n}\{\widehat q_n(\alpha)\},
\qquad i:Z_i=1,\\
\widehat\psi_{i,0,n}(\alpha)
&=\widehat r_n(\alpha)
\bigl[\ind\{Y_i>\widehat q_n(\alpha)\}
-\widehat G_{0,n}\{\widehat q_n(\alpha)\}\bigr],
\qquad i:Z_i=0.
\end{align*}
For $z=0,1$, define the within-arm sample variance
\begin{align*}
\widehat S_{\psi_z\psi_z,n}(\alpha)
=\frac{1}{n_z-1}\sum_{i:Z_i=z}\{\widehat\psi_{i,z,n}(\alpha)\}^2.
\end{align*}
The proposed variance estimator is
\begin{equation}
\widehat\sigma_{R,n}^2(\alpha)
=\frac{\widehat S_{\psi_1\psi_1,n}(\alpha)}{p_n}
+\frac{\widehat S_{\psi_0\psi_0,n}(\alpha)}{1-p_n}.
\label{eq:var-est}
\end{equation}
The corresponding finite-population variance bound is
\begin{align*}
\bar\sigma_{R,n}^2(\alpha)
=\frac{S_{\psi_1\psi_1,n}(\alpha)}{p_n}
+\frac{S_{\psi_0\psi_0,n}(\alpha)}{1-p_n}
=\sigma_{R,n}^2(\alpha)+S_{\delta\delta,n}(\alpha).
\end{align*}

We consider two estimators of $r(\alpha)$, depending on whether numerical outcome values or only pooled ranks are available.
When outcome values are available, we estimate the local densities using symmetric differences of the empirical survival functions. Let $b_n=c_b/\sqrt n$ for some constant $c_b>0$, and define
\begin{align*}
\widehat f_{z,n}\{\widehat q_n(\alpha)\}
=\frac{\widehat G_{z,n}\{\widehat q_n(\alpha)-b_n\}-\widehat G_{z,n}\{\widehat q_n(\alpha)+b_n\}}{2b_n},
\qquad z=0,1.
\end{align*}
The resulting density-ratio slope estimator is given by
\begin{align*}
\widehat r_{d,n}(\alpha)
=\frac{\widehat f_{1,n}\{\widehat q_n(\alpha)\}}
{\widehat f_{0,n}\{\widehat q_n(\alpha)\}}.
\end{align*}
When only pooled ranks are available, we estimate the local ROC slope directly from the empirical ROC curve.
With $b_n=c_b/\sqrt n$ satisfying $\alpha\pm b_n\in(0,1)$, define the rank-based slope estimator as
\begin{align*}
\widehat r_{r,n}(\alpha)
=\frac{\widehat R_n(\alpha+b_n)-\widehat R_n(\alpha-b_n)}{2b_n}.
\end{align*}
The bandwidth $b_n=c_b/\sqrt n$ matches the local scale in Assumption~\ref{a:local}, ensuring that both the outcome-based and rank-based windows remain within the neighborhoods where the local linear approximations hold.
As shown in Section~\ref{sec:app-slopes} of the Supplementary Material, both $\widehat r_{d,n}(\alpha)$ and $\widehat r_{r,n}(\alpha)$ are consistent for $r(\alpha)$, and we write $\widehat r_n(\alpha)$ when no distinction is needed.
We prefer the rank-based estimator $\widehat r_{r,n}(\alpha)$ for two reasons: (i) it relies only on pooled outcome ranks and is therefore invariant to strictly increasing transformations of the outcomes, and (ii) it avoids density estimation.
Moreover, by Proposition~\ref{p:rank}, when $\widehat r_n(\alpha)=\widehat r_{r,n}(\alpha)$, the variance estimator $\widehat\sigma_{R,n}^2(\alpha)$ and the resulting confidence intervals also depend on the outcomes only through their pooled ranks.
However, the consistency of both estimators is established only for fixed $\alpha\in(0,1)$ and is not guaranteed for shrinking tail probabilities of order $O(n^{-1/2})$, which we study in \Cref{sec:tail}.

The following theorem establishes the consistency of the proposed variance estimator for the conservative upper bound and the asymptotic validity of the resulting confidence interval.

\begin{theorem}
\label{t:interval}
Under Assumptions~\ref{a:margins}--\ref{a:limits}, the following hold:
\begin{enumerate}[leftmargin=*,label=(\roman*)]
\item The variance estimator satisfies $\widehat\sigma_{R,n}^2(\alpha)\xrightarrow{p}\bar\sigma_R^2(\alpha)=\lim_{n\to\infty}\bar\sigma_{R,n}^2(\alpha)\ge\sigma_R^2(\alpha)$. Moreover, if $G_1$ is continuous at $q(\alpha)$, then $R_n(\alpha)\to R(\alpha)=G_1\{q(\alpha)\}$ and
\begin{align*}
\bar\sigma_R^2(\alpha)=\frac{R(\alpha)\{1-R(\alpha)\}}{p}+\frac{r(\alpha)^2\alpha(1-\alpha)}{1-p}.
\end{align*}
\item The confidence interval $\widehat R_n(\alpha)\pm z_{1-\eta/2}\widehat\sigma_{R,n}(\alpha)/\sqrt n$
has asymptotic coverage probability of at least $1-\eta$ for $R_n(\alpha)$.
\end{enumerate}
\end{theorem}

\subsection{Covariate adjustment}
\label{sec:adjust}

Covariate adjustment is widely used to improve estimation precision in randomized experiments \citep{lin2013,liding2017}.
For the average treatment effect, subtracting the same function $\varphi(\bm X_i)$ from both potential outcomes leaves the estimand unchanged.
In contrast, directly adjusting potential outcomes is infeasible when only ranks are observed and may alter their relative ordering across units when numerical outcomes are available, thereby changing the ROC curve.
To this end, we instead propose a linear covariate adjustment applied directly to the ROC estimator:
\begin{equation}
\widehat R_{\mathrm{adj},n}(\alpha;\bm\beta)
=\widehat R_n(\alpha)-\bm\beta^\top(\bar{\bm X}_{1,n}-\bar{\bm X}_{0,n}),
\label{eq:adj}
\end{equation}
where $\bar{\bm X}_{z,n}=n_z^{-1}\sum_{i:Z_i=z}\bm X_i$ for $z=0,1$. Under complete randomization, $\E(\bar{\bm X}_{1,n}-\bar{\bm X}_{0,n})=\bm 0$, so the adjustment term has mean zero for any fixed $\bm\beta$.
By \eqref{eq:linear}, for any fixed $\bm\beta$,
\begin{align*}
\widehat R_{\mathrm{adj},n}(\alpha;\bm\beta)-R_n(\alpha)
=L_n(\alpha)-\bm\beta^\top(\bar{\bm X}_{1,n}-\bar{\bm X}_{0,n})+o_P(n^{-1/2}).
\end{align*}
When $\bm S_{\bm X\bm X,n}$ is nonsingular, the coefficient that minimizes the finite-population variance of the leading term is
\begin{equation}
\bm\beta_{*,n}(\alpha)
=\bm S_{\bm X\bm X,n}^{-1}
\bigl\{(1-p_n)\bm S_{\bm X\psi_1,n}(\alpha)
+p_n\bm S_{\bm X\psi_0,n}(\alpha)\bigr\}.
\label{eq:betastar}
\end{equation}
As in regression adjustment for the ATE \citep{lu2025tyranny}, this optimal coefficient is a weighted average of the finite-population regression coefficients of $\psi_{i,1,n}(\alpha)$ and $\psi_{i,0,n}(\alpha)$ on $\bm X_i$, with weights $1-p_n$ and $p_n$, respectively.
Define
\begin{align*}
\psi_{i,z,\mathrm{adj},n}(\alpha)
&=\psi_{i,z,n}(\alpha)-\bm\beta_{*,n}(\alpha)^\top(\bm X_i-\bar{\bm X}_n),
\qquad z=0,1,\\
L_{n,\mathrm{adj}}(\alpha)
&=L_n(\alpha)-\bm\beta_{*,n}(\alpha)^\top(\bar{\bm X}_{1,n}-\bar{\bm X}_{0,n})
=\bar\psi_{1,\mathrm{adj},n}(\alpha)-\bar\psi_{0,\mathrm{adj},n}(\alpha),
\end{align*}
where $\bar\psi_{z,\mathrm{adj},n}(\alpha)=n_z^{-1}\sum_{i:Z_i=z}\psi_{i,z,\mathrm{adj},n}(\alpha)$.

Let $S_{\psi_z\psi_z,\mathrm{adj},n}(\alpha)$ denote the finite-population variance of $\psi_{i,z,\mathrm{adj},n}(\alpha)$.
By Proposition~\ref{p:projection} in the Supplementary Material, the randomization variance of $\sqrt n L_{n,\mathrm{adj}}(\alpha)$ is
\begin{align*}
\sigma^2_{R,\mathrm{adj},n}(\alpha)
&=\frac{S_{\psi_1\psi_1,\mathrm{adj},n}(\alpha)}{p_n}
+\frac{S_{\psi_0\psi_0,\mathrm{adj},n}(\alpha)}{1-p_n}
-S_{\delta\delta,n}(\alpha)\\
&=\sigma_{R,n}^2(\alpha)
-\frac{\bm\beta_{*,n}(\alpha)^\top\bm S_{\bm X\bm X,n}\bm\beta_{*,n}(\alpha)}{p_n(1-p_n)}
\le\sigma_{R,n}^2(\alpha).
\end{align*}

\begin{assumption}
\label{a:adjust}
The dimension $d$ is fixed, $\bm S_{\bm X\bm X,n}$ converges to a positive-definite matrix, and $\bm S_{\bm X\psi_z,n}(\alpha)$ converges to a finite vector for $z=0,1$. Moreover,
$n^{-1}\max_{1\le i\le n}\|\bm X_i-\bar{\bm X}_n\|^2\to0$, and $\sigma^2_{R,\mathrm{adj},n}(\alpha)\to\sigma^2_{R,\mathrm{adj}}(\alpha)>0$.
\end{assumption}

Assumption~\ref{a:adjust} imposes standard finite-population regularity conditions for regression adjustment \citep{lin2013,liding2017,chen2026design}, applied here to $\psi_{i,z,n}(\alpha)$.

To implement the adjustment and variance estimation, we estimate $\bm\beta_{*,n}(\alpha)$ by
\begin{equation}
\widehat{\bm\beta}_n(\alpha)
=\bm S_{\bm X\bm X,n}^{-1}
\bigl\{(1-p_n)\widehat{\bm S}_{\bm X\psi_1,n}(\alpha)
+p_n\widehat{\bm S}_{\bm X\psi_0,n}(\alpha)\bigr\},
\label{eq:betahat}
\end{equation}
where
\begin{align*}
\widehat{\bm S}_{\bm X\psi_1,n}(\alpha)&=\frac{1}{(n-1)p_n}\sum_{i:Z_i=1}(\bm X_i-\bar{\bm X}_n)\ind\{Y_i>\widehat q_n(\alpha)\},\\
\widehat{\bm S}_{\bm X\psi_0,n}(\alpha)&=\frac{\widehat r_n(\alpha)}{(n-1)(1-p_n)}\sum_{i:Z_i=0}(\bm X_i-\bar{\bm X}_n)\ind\{Y_i>\widehat q_n(\alpha)\},
\end{align*}
and $\widehat r_n(\alpha)$ denotes either slope estimator $\widehat r_{d,n}(\alpha)$ or $\widehat r_{r,n}(\alpha)$ from Section~\ref{sec:interval}.
The estimated adjusted linearization terms are then given by
\begin{align*}
\widehat\psi_{i,z,\mathrm{adj},n}(\alpha)
=\widehat\psi_{i,z,n}(\alpha)-\widehat{\bm\beta}_n(\alpha)^\top(\bm X_i-\bar{\bm X}_n),
\qquad z=0,1,
\end{align*}
with within-arm sample variances
\begin{align*}
\widehat S_{\psi_z\psi_z,\mathrm{adj},n}(\alpha)
=\frac{1}{n_z-1}\sum_{i:Z_i=z}
\bigl\{\widehat\psi_{i,z,\mathrm{adj},n}(\alpha)
-\widehat{\bar\psi}_{z,\mathrm{adj},n}(\alpha)\bigr\}^2,
\end{align*}
where $\widehat{\bar\psi}_{z,\mathrm{adj},n}(\alpha)
=n_z^{-1}\sum_{i:Z_i=z}\widehat\psi_{i,z,\mathrm{adj},n}(\alpha)$.
The adjusted variance estimator is given by
\begin{align*}
\widehat\sigma_{R,\mathrm{adj},n}^2(\alpha)
=\frac{\widehat S_{\psi_1\psi_1,\mathrm{adj},n}(\alpha)}{p_n}
+\frac{\widehat S_{\psi_0\psi_0,\mathrm{adj},n}(\alpha)}{1-p_n}.
\end{align*}
When the rank-based slope estimator $\widehat r_{r,n}(\alpha)$ is used, the quantities $\widehat{\bm\beta}_n(\alpha)$, $\widehat R_{\mathrm{adj},n}\{\alpha;\widehat{\bm\beta}_n(\alpha)\}$, and $\widehat\sigma_{R,\mathrm{adj},n}^2(\alpha)$ depend only on the pooled ranks, thereby extending Proposition~\ref{p:rank} to covariate-adjusted inference.

\begin{theorem}
\label{t:adjust}
Under Assumptions~\ref{a:margins}--\ref{a:adjust}, the following hold:
\begin{enumerate}[leftmargin=*,label=(\roman*)]
\item The estimated adjustment coefficients satisfy $\|\widehat{\bm\beta}_n(\alpha)-\bm\beta_{*,n}(\alpha)\|=o_P(1)$, and
\begin{align*}
\sqrt n\bigl[\widehat R_{\mathrm{adj},n}\{\alpha;\widehat{\bm\beta}_n(\alpha)\}-R_n(\alpha)\bigr]
\xrightarrow{d}\mathcal N\{0,\sigma^2_{R,\mathrm{adj}}(\alpha)\},\qquad \text{with}\qquad \sigma^2_{R,\mathrm{adj}}(\alpha)\le\sigma_R^2(\alpha).
\end{align*}
\item The variance estimator
satisfies $\widehat\sigma_{R,\mathrm{adj},n}^2(\alpha)\xrightarrow{p}\bar\sigma^2_{R,\mathrm{adj}}(\alpha)\ge\sigma^2_{R,\mathrm{adj}}(\alpha)$, where
\begin{align*}
\begin{aligned}
\bar\sigma^2_{R,\mathrm{adj}}(\alpha)=\lim_{n\to\infty}\bar\sigma^2_{R,\mathrm{adj},n}(\alpha),\quad 
\bar\sigma^2_{R,\mathrm{adj},n}(\alpha)
=\frac{S_{\psi_1\psi_1,\mathrm{adj},n}(\alpha)}{p_n}
+\frac{S_{\psi_0\psi_0,\mathrm{adj},n}(\alpha)}{1-p_n}.
\end{aligned}
\end{align*}
Consequently, the confidence interval $\widehat R_{\mathrm{adj},n}\{\alpha;\widehat{\bm\beta}_n(\alpha)\}\pm z_{1-\eta/2}\widehat\sigma_{R,\mathrm{adj},n}(\alpha)/\sqrt n$ has asymptotic coverage at least $1-\eta$.
\end{enumerate}
\end{theorem}

Theorem~\ref{t:adjust} establishes the asymptotic normality of the adjusted estimator and guarantees that covariate adjustment does not increase its asymptotic variance.
Moreover, because the adjustment leaves the unidentifiable variance component $S_{\delta\delta,n}(\alpha)$ unchanged, it yields asymptotically no wider confidence intervals.

\section{Simultaneous inference}
\label{sec:simultaneous}

In many applications, interest lies in comparing treatment and control distributions across a range of thresholds rather than at a single fixed threshold.
In this section, we construct simultaneous confidence bands over a fixed interval $\mathcal A=[\varepsilon,1-\varepsilon]$, where $\varepsilon\in(0,1/2)$.
To this end, we strengthen the pointwise regularity conditions to hold uniformly over $\mathcal A$ and establish functional convergence of the ROC estimator.
For $\alpha,\alpha'\in\mathcal A$, define the finite-population covariance kernel
\begin{align*}
K_n(\alpha,\alpha')
=\frac{1}{n-1}\sum_{i=1}^n\biggl\{
\frac{\psi_{i,1,n}(\alpha)\psi_{i,1,n}(\alpha')}{p_n}
+\frac{\psi_{i,0,n}(\alpha)\psi_{i,0,n}(\alpha')}{1-p_n}
-\delta_{i,n}(\alpha)\delta_{i,n}(\alpha')
\biggr\},
\end{align*}
where $\delta_{i,n}(\alpha)=\psi_{i,1,n}(\alpha)-\psi_{i,0,n}(\alpha)$, so that $K_n(\alpha,\alpha)=\sigma_{R,n}^2(\alpha)$.

\begin{assumption}
\label{a:uniform}
Uniformly over $\alpha\in\mathcal A$, $q_n(\alpha)\to q(\alpha)$, where $G_0\{q(\alpha)\}=\alpha$.
The limiting survival functions $G_0$ and $G_1$ are continuous at $q(\alpha)$ for every $\alpha\in\mathcal A$.
For $z=0,1$ and every fixed $M>0$,
\begin{align*}
\sup_{\alpha\in\mathcal A}\sup_{|t|\le M}
\left|
\sqrt n\left[G_{z,n}\{q_n(\alpha)+t/\sqrt n\}-G_{z,n}\{q_n(\alpha)\}\right]
+t f_z\{q(\alpha)\}
\right|\to0.
\end{align*}
For $z=0,1$, the function $\alpha\mapsto f_z\{q(\alpha)\}$ is continuous and bounded away from zero and infinity on $\mathcal A$.
Moreover, $K_n(\alpha,\alpha')$ converges uniformly to a covariance kernel $K(\alpha,\alpha')$ on $\mathcal A\times\mathcal A$, with $\inf_{\alpha\in\mathcal A}K(\alpha,\alpha)>0$.
\end{assumption}

\begin{theorem}
\label{t:fclt}
Under Assumptions~\ref{a:margins}, \ref{a:limits} and~\ref{a:uniform},
\begin{align*}
\sqrt n\{\widehat R_n-R_n\}\rightsquigarrow\mathbb G_R
\qquad\text{in }\ell^\infty(\mathcal A),
\end{align*}
where $\mathbb G_R$ is a mean-zero Gaussian process with covariance kernel $K(\alpha,\alpha')$, and $\ell^\infty(\mathcal A)$ denotes the space of bounded real-valued functions on $\mathcal A$.
\end{theorem}

Theorem~\ref{t:fclt} provides the basis for simultaneous inference over $\mathcal A$.
Because $K$ is not identifiable, we consider the conservative covariance kernel
\begin{align*}
\bar K_n(\alpha,\alpha')
=\frac1{n-1}\sum_{i=1}^n\left\{
\frac{\psi_{i,1,n}(\alpha)\psi_{i,1,n}(\alpha')}{p_n}
+\frac{\psi_{i,0,n}(\alpha)\psi_{i,0,n}(\alpha')}{1-p_n}
\right\}.
\end{align*}
Under Assumptions~\ref{a:margins}, \ref{a:limits} and~\ref{a:uniform}, its uniform limit has the closed form
\begin{align*}
\bar K(\alpha,\alpha')
=\frac{R(\alpha\wedge\alpha')\{1-R(\alpha\vee\alpha')\}}{p}
+\frac{r(\alpha)r(\alpha')(\alpha\wedge\alpha')(1-\alpha\vee\alpha')}{1-p},
\end{align*}
where $\alpha\wedge\alpha'=\min(\alpha,\alpha')$ and $\alpha\vee\alpha'=\max(\alpha,\alpha')$. This two-Brownian-bridge covariance structure directly generalizes the explicit conservative variance formula in Theorem~\ref{t:interval}(i), with $\bar K(\alpha,\alpha)=\bar\sigma_R^2(\alpha)$.
We estimate $\bar K$ by
\begin{align*}
\widehat K_n(\alpha,\alpha')
=\frac1{p_n(n_1-1)}
\sum_{Z_i=1}
\widehat\psi_{i,1,n}(\alpha)
\widehat\psi_{i,1,n}(\alpha')+\frac1{(1-p_n)(n_0-1)}
\sum_{Z_i=0}
\widehat\psi_{i,0,n}(\alpha)
\widehat\psi_{i,0,n}(\alpha').
\end{align*}
In particular, $\widehat K_n(\alpha,\alpha)=\widehat\sigma_{R,n}^2(\alpha)$, recovering the pointwise variance estimator.

We construct simultaneous confidence bands by simulating Gaussian vectors with covariance matrix estimated from $\widehat K_n$ on a grid over $\mathcal A$.
The procedure is summarized in Algorithm~\ref{alg:gaussian-band}.

\begin{algorithm}[H]
\caption{Simultaneous confidence band construction}
\label{alg:gaussian-band}
\begin{algorithmic}[1]
\REQUIRE $\varepsilon\in(0,1/2)$, observed data, a grid $\varepsilon=\alpha_1<\cdots<\alpha_m=1-\varepsilon$, significance level $\eta\in(0,1)$, and number of simulations $B$.
\STATE Compute $\widehat R_n(\alpha_j)$ and $\widehat\psi_{i,z,n}(\alpha_j)$ for $z=0,1$ at each grid point $\alpha_j$.
\STATE Form $\widehat{\bm\Sigma}_n=\{\widehat K_n(\alpha_j,\alpha_k)\}_{j,k=1}^m$ and set $\widehat s_j=\{\widehat K_n(\alpha_j,\alpha_j)\}^{1/2}$.
\STATE For $b=1,\ldots,B$, independently draw $\bm\zeta^{(b)}\sim\mathcal N_m(\bm0,\widehat{\bm\Sigma}_n)$ and compute
$T^{(b)}=\max_{1\le j\le m}|\zeta_j^{(b)}|/\widehat s_j$.
\STATE Let $\widetilde c_{1-\eta,n}$ be the empirical $(1-\eta)$-quantile of $T^{(1)},\ldots,T^{(B)}$.
\ENSURE The band $\widehat R_n(\alpha_j)\pm\widetilde c_{1-\eta,n}\widehat s_j/\sqrt n$ at each grid point.
\end{algorithmic}
\end{algorithm}

\begin{corollary}
\label{c:band}
Under Assumptions~\ref{a:margins}, \ref{a:limits} and~\ref{a:uniform}, let $\widetilde c_{1-\eta,n}$ be obtained from Algorithm~\ref{alg:gaussian-band}, with $B\to\infty$ and $\max_{j<m}(\alpha_{j+1}-\alpha_j)\to0$.
Then
\begin{align*}
\liminf_{n\to\infty}\Pr\Bigl\{
R_n(\alpha)\in
\widehat R_n(\alpha)\pm
\widetilde c_{1-\eta,n}
\bigl\{\widehat K_n(\alpha,\alpha)/n\bigr\}^{1/2}
\ \text{for all }\alpha\in\mathcal A
\Bigr\}\ge1-\eta.
\end{align*}
\end{corollary}

Corollary~\ref{c:band} establishes asymptotically conservative simultaneous coverage of the ROC curve over $\mathcal A$. In particular, the band can be directly used to test the uniform null hypothesis of no distributional treatment effect, $H_0\colon R_n(\alpha)=\alpha$ for all $\alpha\in\mathcal A$, by checking whether the diagonal lies entirely within the band, as illustrated in Section~\ref{sec:real-data} and Figure~\ref{fig:real-data-roc}.
The covariate adjustment methods of \Cref{sec:adjust} also extend to simultaneous inference; see Section~\ref{sec:app-scb} of the Supplementary Material for details. Moreover, as in Proposition~\ref{p:rank}, when the rank-based slope estimator $\widehat r_{r,n}(\alpha)$ is used, Algorithm~\ref{alg:gaussian-band} and the resulting simultaneous confidence bands (including the covariate-adjusted version in Section~\ref{sec:app-scb}) depend on the outcomes only through their pooled ranks.

\section{Tail inference}
\label{sec:tail}

The preceding sections develop pointwise and simultaneous inference for the ROC curve at fixed control exceedance probabilities $\alpha\in(0,1)$.
In many applications, however, interest lies in extreme upper-tail outcomes, such as whether an intervention increases the proportion of units exceeding the top $1\%$ or $0.1\%$ of control outcomes. This motivates studying shrinking exceedance probabilities $\alpha_n\to0$, namely high quantiles \citep{DekkersdeHaan1989,dehaanferreira2006,Zhang2018}.
The preceding results do not directly apply to this setting.
In particular, the local slope estimators with bandwidth $b_n=c_b/\sqrt n$ are unsuitable for tail probabilities of order $O(n^{-1/2})$.
We therefore develop inference for the ROC curve at shrinking control exceedance probabilities.

For fixed $\alpha$, Assumption~\ref{a:local} imposes local linearity around $q(\alpha)$ with positive limiting densities.
When $\alpha_n\to0$, however, the control quantile $q_n(\alpha_n)$ may diverge, while the densities $f_z\{q_n(\alpha_n)\}$ may vanish.
To accommodate this behavior, we formulate regularity conditions directly on the control-percentile scale.
Specifically, define $V_i(z)=G_{0,n}\{Y_i(z)\}\in[0,1]$ and
\begin{align*}
H_{z,n}(v)=\frac1n\sum_{i=1}^n\ind\{V_i(z)\le v\},
\qquad z=0,1.
\end{align*}
Here $V_i(z)$ is the proportion of control potential outcomes strictly exceeding $Y_i(z)$.
Since $G_{0,n}$ is nonincreasing, larger outcomes correspond to smaller values of $V_i(z)$, with extreme upper-tail outcomes mapped to values near zero.
On this scale, we require $H_{0,n}(v)$ and $H_{1,n}(v)$ to be approximated by $v$ and the limiting ROC curve $R(v)$, respectively, near zero.

\begin{assumption}
\label{a:tail}
For some fixed $v_0\in(0,1)$ and a function $R:(0,v_0]\to(0,1]$ with derivative $r=R'$, the following conditions hold:
\begin{enumerate}[leftmargin=*,label=(\roman*)]
\item $R$ is continuous and increasing on $(0,v_0]$ with $\lim_{v\to0^+}R(v)=0$, $H_{0,n}(c\alpha_n)/(c\alpha_n)\to 1$ and $H_{1,n}(c\alpha_n)/R(c\alpha_n)\to 1$ for every fixed $c>0$, and, for every fixed $M>0$,
\begin{align*}
&\sup_{|t|\le M}
\bigl|H_{0,n}\bigl(\alpha_n+t\sqrt{\alpha_n/n}\bigr)
-\alpha_n-t\sqrt{\alpha_n/n}\bigr|
=o\bigl(\sqrt{\alpha_n/n}\bigr),\\
&\sup_{|t|\le M}
\bigl|H_{1,n}\bigl(\alpha_n+t\sqrt{\alpha_n/n}\bigr)
-H_{1,n}(\alpha_n)
-r(\alpha_n)t\sqrt{\alpha_n/n}\bigr|\\
&\qquad=o\bigl\{\sqrt{R(\alpha_n)/n}+r(\alpha_n)\sqrt{\alpha_n/n}\bigr\}.
\end{align*}

\item $R$ is continuously differentiable on $(0,v_0]$ with $r(v)=R'(v)>0$, $r$ is locally absolutely continuous on $(0,v_0]$, and for some $\gamma>0$ and $C>0$,
\begin{align*}
\lim_{v\to0^+}\frac{v r(v)}{R(v)}=\gamma,
\qquad
|r'(v)|\le C\frac{r(v)}v
\quad\text{for almost every }v\in(0,v_0].
\end{align*}

\item The sequence $\alpha_n\in(0,1)$ satisfies $\alpha_n\to0$, $n\alpha_n\to\infty$, and $nR(\alpha_n)\to\infty$.
\end{enumerate}
\end{assumption}

Assumption~\ref{a:tail}(i) extends Assumptions~\ref{a:margins} and~\ref{a:local} to shrinking control exceedance probabilities at the corresponding tail scales, requiring relative convergence of $H_{z,n}$ at scale $\alpha_n$ and local linear approximations around $\alpha_n$ at scale $\sqrt{\alpha_n/n}$.
Assumption~\ref{a:tail}(ii) imposes regular variation of $R$ at zero with index $\gamma>0$ \citep{bingham1987,dehaanferreira2006}, together with a derivative bound controlling the curvature of $R$ near zero \citep{csorgorevesz1978}.
As $\alpha\to0^+$, it gives $r(\alpha)=\{\gamma+o(1)\}R(\alpha)/\alpha$ and implies $R(t\alpha)/R(\alpha)\to t^\gamma$ for every $t>0$.
Thus, $\gamma$ characterizes the relative upper-tail behavior of the two potential outcome distributions: when $\gamma<1$, the treatment distribution has a heavier upper tail, with $R(\alpha)/\alpha\to\infty$ and $r(\alpha)\to\infty$; when $\gamma=1$, the two upper tails decay at comparable rates up to a slowly varying factor; and when $\gamma>1$, the treatment upper tail is lighter, with $R(\alpha)/\alpha\to0$ and $r(\alpha)\to0$.
Together with $p_n\to p\in(0,1)$, Assumption~\ref{a:tail}(iii) ensures that the expected numbers of tail observations in both treatment arms diverge.

As an example, suppose $F_0$ and $F_1$ are continuous distributions whose survival functions $G_z=1-F_z$ are twice continuously differentiable in the upper tail, with $G_z'<0$, and satisfy $G_z(y)=a_z y^{-\nu_z}\{1+o(1)\}$, $G_z'(y)=-\nu_z a_z y^{-\nu_z-1}\{1+o(1)\}$, and $G_z''(y)=\nu_z(\nu_z+1)a_z y^{-\nu_z-2}\{1+o(1)\}$ as $y\to\infty$, where $a_z,\nu_z>0$, as holds for Pareto and Student-$t$ distributions.
With $\gamma=\nu_1/\nu_0$ and $c=a_1a_0^{-\gamma}$, we have $R(v)=cv^\gamma\{1+o(1)\}$, $r(v)=c\gamma v^{\gamma-1}\{1+o(1)\}$, and $vr'(v)/r(v)\to\gamma-1$ as $v\to0^+$, verifying Assumption~\ref{a:tail}(ii); by invariance under common strictly increasing transformations, light-tailed families such as $\mathrm{Exp}(\lambda_z)$, for which $R(v)=v^{\lambda_1/\lambda_0}$, are also covered.
Moreover, if $\alpha_n\to0$, $n\alpha_n\to\infty$, and $n\alpha_n^\gamma\to\infty$, Assumptions~\ref{a:tail}(i) and~(iii) hold for finite populations without control ties satisfying $\sup_{y\in\mathbb R}|G_{z,n}(y)-G_z(y)|=O(n^{-1})$ ($z=0,1$), and hold in probability when $\{Y_i(1),Y_i(0)\}_{i=1}^n$ are drawn independently from a joint distribution with marginals $(F_1,F_0)$.

The linearization terms $\psi_{i,z,n}(\alpha_n)$ and $\delta_{i,n}(\alpha_n)$, together with the variances $\sigma_{R,n}^2(\alpha_n)$ and $\bar\sigma_{R,n}^2(\alpha_n)$, are defined as in Sections~\ref{sec:normal} and~\ref{sec:interval}, with $\alpha$ replaced by $\alpha_n$.
Constructing the variance estimator $\widehat\sigma_{R,n}^2(\alpha_n)$ requires an estimator of $r(\alpha_n)$ satisfying $\widehat r_n(\alpha_n)/r(\alpha_n)\xrightarrow{p}1$. The consistency results for the local-difference estimators $\widehat r_{d,n}$ and $\widehat r_{r,n}$ do not automatically extend to $\alpha_n=O(n^{-1/2})$, since the rank-based window $[\alpha_n-b_n,\alpha_n+b_n]$ is no longer negligible relative to $\alpha_n$ and may exit $(0,1)$, and the outcome-based density estimates may fail to achieve relative consistency.
We instead construct a slope estimator using Assumption~\ref{a:tail}(ii), which implies
\begin{align*}
\lim_{\alpha\to0^+}
\frac{\mathrm d\log R(\alpha)}{\mathrm d\log\alpha}
=
\lim_{\alpha\to0^+}
\frac{\alpha R'(\alpha)}{R(\alpha)}=
\lim_{\alpha\to0^+}
\frac{\alpha r(\alpha)}{R(\alpha)}
=\gamma.
\end{align*}
By the mean value theorem,
there exists $\xi_n\in(\alpha_n/2,\alpha_n)$ such that
\begin{align*}
\frac{\log R(\alpha_n)-\log R(\alpha_n/2)}{\log 2}
&=\frac{\xi_n r(\xi_n)}{R(\xi_n)}
\to\gamma.
\end{align*}
We can therefore estimate $\gamma$ via
\begin{align*}
\widehat\gamma_n
=
\frac{\log\widehat R_n(\alpha_n)
-\log\widehat R_n(\alpha_n/2)}
{\log 2},
\end{align*}
which is nonnegative since $\widehat R_n$ is nondecreasing.
Assumption~\ref{a:tail}(ii) gives $r(\alpha_n)=\{\gamma+o(1)\}R(\alpha_n)/\alpha_n$, so estimating $\gamma$ and $R(\alpha_n)$ by their empirical counterparts gives the tail slope estimator
\begin{equation}
\widehat r_{t,n}(\alpha_n)
=\widehat\gamma_n\widehat R_n(\alpha_n)/\alpha_n,
\label{eq:tail-slope}
\end{equation}
which satisfies $\widehat r_{t,n}(\alpha_n)/r(\alpha_n)\xrightarrow{p}1$, as shown in Section~\ref{sec:app-slopes} of the Supplementary Material.

If either $\widehat R_n(\alpha_n)$ or $\widehat R_n(\alpha_n/2)$ is zero, we set $\widehat\gamma_n=1$, and if $\widehat R_n(\alpha_n)=0$, we further set $\widehat r_{t,n}(\alpha_n)=1$. Under Assumption~\ref{a:tail}, this event has probability tending to zero, so the convention does not affect the asymptotic results. Substituting $\widehat r_n(\alpha_n)=\widehat r_{t,n}(\alpha_n)$ into the linearization terms and conservative variance estimator of Section~\ref{sec:interval} gives $\widehat\psi_{i,z,n}(\alpha_n)$ and $\widehat\sigma_{R,n}^2(\alpha_n)$.

Both the tail slope estimator and the exceedance indicators at $\widehat q_n(\alpha_n)$ can be computed from pooled ranks.
Consequently, the tail variance estimator and confidence interval are also rank-based and invariant to strictly increasing transformations of the outcomes.

\begin{theorem}
\label{t:tail-clt}
Under Assumption~\ref{a:tail},
the following hold:
\begin{enumerate}[leftmargin=*,label=(\roman*)]
\item $\sigma_{R,n}^2(\alpha_n)\asymp\bar\sigma_{R,n}^2(\alpha_n)\asymp R(\alpha_n)+r(\alpha_n)^2\alpha_n$, and $\bar\sigma_{R,n}^2(\alpha_n)-\sigma_{R,n}^2(\alpha_n)=S_{\delta\delta,n}(\alpha_n)\ge0$.
\item $\sqrt n\{\widehat R_n(\alpha_n)-R_n(\alpha_n)\}/\sigma_{R,n}(\alpha_n)\xrightarrow{d}\mathcal N(0,1)$.
\item $\widehat\sigma_{R,n}^2(\alpha_n)/\bar\sigma_{R,n}^2(\alpha_n)\xrightarrow{p}1$, 
and the interval
$\widehat R_n(\alpha_n)\pm z_{1-\eta/2}\widehat\sigma_{R,n}(\alpha_n)/\sqrt n$
has asymptotic coverage at least $1-\eta$.
\end{enumerate}
\end{theorem}

Theorem~\ref{t:tail-clt} establishes asymptotic normality and valid conservative inference for the ROC curve at shrinking exceedance probabilities.
The two terms $R(\alpha_n)$ and $r(\alpha_n)^2\alpha_n$ in the variance order represent treatment-arm fluctuations and uncertainty in estimating the control tail threshold, respectively.
Their relative importance depends on the tail index $\gamma$: when $\gamma<1$ ($R(\alpha_n)\gg\alpha_n$), control threshold uncertainty dominates; when $\gamma=1$ with $R(\alpha_n)\asymp\alpha_n$, both contribute at the same order; and when $\gamma>1$ ($R(\alpha_n)\ll\alpha_n$), treatment-arm fluctuations dominate.
Because the slope enters the variance only through the control-threshold term, relative consistency of the slope estimator is essential for $\gamma\le 1$ (whenever $R(\alpha_n)\gtrsim\alpha_n$).
The results can also be extended to incorporate covariate adjustment, as detailed in Section~\ref{sec:tail-adjust} of the Supplementary Material.

Table~\ref{tab:slope-domains} compares the applicable regimes of the three slope estimators.
For fixed $\alpha\in(0,1)$, the rank-based estimator $\widehat r_{r,n}$ is preferred, while $\widehat r_{d,n}$ is also consistent when numerical outcomes are available; the tail-based estimator $\widehat r_{t,n}$ is generally biased.
For shrinking $\alpha_n\to0$, the tail estimator is relatively consistent under Assumption~\ref{a:tail}, including the regime $n^{-1}\ll\alpha_n\lesssim n^{-1/2}$ with $nR(\alpha_n)\to\infty$.
The rank-based and density-ratio estimators are relatively consistent under the additional conditions in Proposition~\ref{p:slope-tail}(ii) and~(iii), respectively.
See Section~\ref{sec:app-slopes} of the Supplementary Material for details.

\begin{table}[!htbp]
\centering
\caption{Applicable regimes of the three slope estimators.}
\label{tab:slope-domains}
\footnotesize
\setlength{\tabcolsep}{4pt}
\begin{tabular}{llll}
\toprule
Estimator & Fixed $\alpha$ & Overlap ($n^{-1/2}\ll\alpha_n\to0$) & Tail ($\alpha_n\lesssim n^{-1/2}$) \\
\midrule
Density-ratio $\widehat r_{d,n}$ & Consistent & Consistent under Prop.~\ref{p:slope-tail}(iii) & Not guaranteed \\
Rank-based $\widehat r_{r,n}$ & Consistent (preferred) & Consistent under Prop.~\ref{p:slope-tail}(ii) & Not guaranteed \\
Tail $\widehat r_{t,n}$ & Generally biased & Consistent under Prop.~\ref{p:slope-tail}(i) & Consistent \\
\bottomrule
\end{tabular}
\end{table}

\section{Simulation}
\label{sec:simulation}

In this section, we evaluate the finite-sample performance of the proposed methods under complete randomization.
For each setting, we generate a finite population of size $n\in\{1000,2000,5000,10000\}$ and draw 2000 independent treatment assignments with treatment proportion $p_n\in\{0.3,0.5\}$, holding the finite population fixed across randomizations.
The potential outcomes are generated as
\begin{align*}
Y_i(0)=\bm\beta_0^\top\bm X_i+\epsilon_i(0),\quad
Y_i(1)=\theta+\bm\beta_1^\top\bm X_i+\epsilon_i(1),
\end{align*}
where $\bm X_i\in\mathbb R^2$ has independent $t_3$ components, $\bm\beta_0=(2,2)^\top$, $\bm\beta_1=(1,3)^\top$, $\theta=1$, and $\epsilon_i(0)$ and $\epsilon_i(1)$ are independent $t_3$ variables.

For pointwise and simultaneous inference, we use the rank-based slope estimator $\widehat r_{r,n}(\alpha)$ with bandwidth $b_n=n^{-1/2}$.
For tail inference, we compare the tail estimator $\widehat r_{t,n}(\alpha_n)$ with $\widehat r_{r,n}(\alpha_n)$ and the density-ratio estimator $\widehat r_{d,n}(\alpha_n)$.

\begin{table}[!t]
\centering
\caption{Coverage probabilities and average lengths of 95\% pointwise confidence intervals.}
\label{tab:sim-pointwise}
\setlength{\tabcolsep}{1.5pt}
\begin{tabular}{llcccccccccccc}
\toprule
& & \multicolumn{2}{c}{$\alpha=0.05$} & \multicolumn{2}{c}{$\alpha=0.10$} & \multicolumn{2}{c}{$\alpha=0.25$} & \multicolumn{2}{c}{$\alpha=0.50$} & \multicolumn{2}{c}{$\alpha=0.75$} & \multicolumn{2}{c}{Length} \\
\cmidrule(lr){3-4}\cmidrule(lr){5-6}\cmidrule(lr){7-8}\cmidrule(lr){9-10}\cmidrule(lr){11-12}\cmidrule(lr){13-14}
$n$ & $p_n$ & Unadj. & Adj. & Unadj. & Adj. & Unadj. & Adj. & Unadj. & Adj. & Unadj. & Adj. & Unadj. & Adj. \\
\midrule
\multirow{2}{*}{1000} & 0.3 & 0.947 & 0.958 & 0.955 & 0.972 & 0.971 & 0.984 & 0.961 & 0.974 & 0.965 & 0.967 & 0.111 & 0.088 \\
 & 0.5 & 0.930 & 0.920 & 0.974 & 0.978 & 0.962 & 0.980 & 0.964 & 0.971 & 0.974 & 0.971 & 0.102 & 0.080 \\
\addlinespace[3pt]
\multirow{2}{*}{2000} & 0.3 & 0.956 & 0.964 & 0.951 & 0.960 & 0.967 & 0.981 & 0.967 & 0.983 & 0.972 & 0.974 & 0.075 & 0.061 \\
 & 0.5 & 0.976 & 0.981 & 0.961 & 0.970 & 0.975 & 0.986 & 0.967 & 0.986 & 0.977 & 0.982 & 0.069 & 0.056 \\
\addlinespace[3pt]
\multirow{2}{*}{5000} & 0.3 & 0.963 & 0.974 & 0.974 & 0.980 & 0.973 & 0.981 & 0.960 & 0.981 & 0.964 & 0.975 & 0.048 & 0.039 \\
 & 0.5 & 0.957 & 0.967 & 0.964 & 0.976 & 0.975 & 0.986 & 0.970 & 0.982 & 0.969 & 0.981 & 0.044 & 0.036 \\
\addlinespace[3pt]
\multirow{2}{*}{10000} & 0.3 & 0.978 & 0.984 & 0.971 & 0.979 & 0.965 & 0.979 & 0.965 & 0.975 & 0.962 & 0.977 & 0.034 & 0.028 \\
 & 0.5 & 0.976 & 0.981 & 0.981 & 0.985 & 0.970 & 0.982 & 0.973 & 0.984 & 0.975 & 0.986 & 0.031 & 0.025 \\
\bottomrule
\end{tabular}

\vspace{2pt}
\parbox{\linewidth}{\footnotesize
\textit{Note.} Lengths are averaged over the five values of $\alpha$.
}
\end{table}

\paragraph*{Pointwise inference.} 
Table~\ref{tab:sim-pointwise} reports empirical coverage probabilities and average lengths of 95\% pointwise confidence intervals.
Both unadjusted and adjusted intervals achieve coverage close to or above the nominal level in most settings.
Covariate adjustment reduces the average interval length by 18.2\%--20.9\% while maintaining valid inference.

\begin{figure}[!htbp]
\centering
\includegraphics[width=0.88\textwidth]{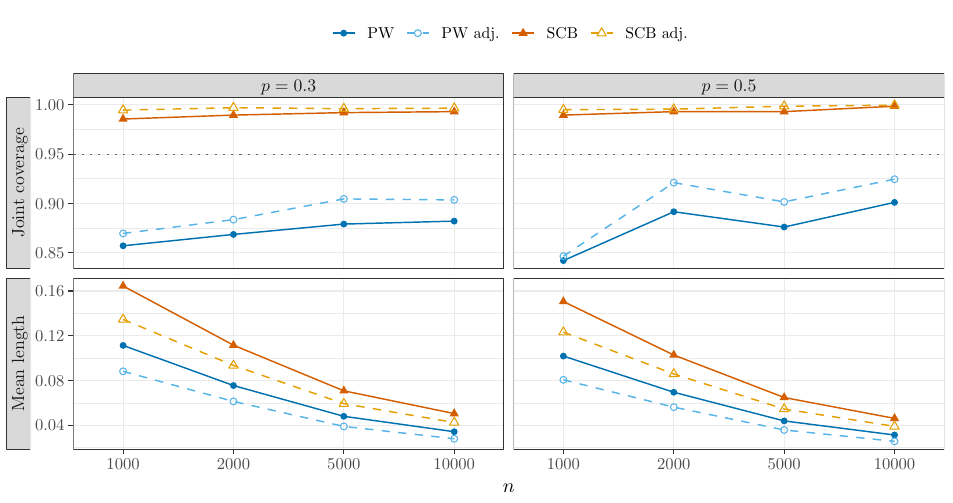}
\caption{Joint coverage probabilities (upper row) and average interval lengths (lower row) for pointwise intervals (PW) and simultaneous confidence bands (SCB), with and without covariate adjustment. The dotted line indicates the nominal 95\% coverage level.}
\label{fig:sim-joint}
\end{figure}

\paragraph*{Simultaneous inference.}
Although the pointwise intervals achieve approximately nominal coverage at each $\alpha$, they do not guarantee joint coverage across multiple values of $\alpha$.
Figure~\ref{fig:sim-joint} shows that their joint coverage at the same five values of $\alpha$ as in Table~\ref{tab:sim-pointwise} falls below the nominal 95\% level.
To address this issue, we construct simultaneous confidence bands using Algorithm~\ref{alg:gaussian-band} on the grid $\{0.05,0.075,\ldots,0.95\}$, with bandwidth $b_n=n^{-1/2}$ and $B=2000$ Gaussian draws.
As shown in Figure~\ref{fig:sim-joint}, the resulting bands achieve conservative joint coverage, while covariate adjustment reduces their average length by 15.5\%--18.2\% while maintaining valid inference.

\begin{table}[!t]
\centering
\caption{Coverage probabilities and average lengths of 95\% confidence intervals under three choices of $\alpha_n$ with $p_n=0.5$.}
\label{tab:sim-tail}
\footnotesize
\setlength{\tabcolsep}{3.5pt}
\begin{tabular}{llcccccccccc}
\toprule
& & & & \multicolumn{2}{c}{Density-ratio} & \multicolumn{2}{c}{Rank-based} & \multicolumn{2}{c}{Tail} & \multicolumn{2}{c}{Length} \\
\cmidrule(lr){5-6}\cmidrule(lr){7-8}\cmidrule(lr){9-10}\cmidrule(lr){11-12}
Choice of $\alpha_n$ & $n$ & $\alpha_n$ & $R_n(\alpha_n)$ & Unadj. & Adj. & Unadj. & Adj. & Unadj. & Adj. & Unadj. & Adj. \\
\midrule
\multirow{3}{*}{$\alpha_n=n^{-1/3}$}
 & 2000 & 0.079 & 0.111 & 0.945 & 0.955 & 0.976 & 0.983 & 0.969 & 0.977 & 0.059 & 0.050 \\
 & 5000 & 0.058 & 0.093 & 0.927 & 0.932 & 0.975 & 0.981 & 0.979 & 0.984 & 0.040 & 0.034 \\
 & 10000 & 0.046 & 0.073 & 0.937 & 0.950 & 0.975 & 0.983 & 0.975 & 0.981 & 0.023 & 0.020 \\
\addlinespace[2pt]
\multirow{3}{*}{$\alpha_n=1.5n^{-1/2}$}
 & 2000 & 0.034 & 0.049 & 0.872 & 0.889 & 0.970 & 0.973 & 0.967 & 0.971 & 0.044 & 0.039 \\
 & 5000 & 0.021 & 0.030 & 0.922 & 0.929 & 0.983 & 0.989 & 0.965 & 0.978 & 0.019 & 0.017 \\
 & 10000 & 0.015 & 0.024 & 0.854 & 0.861 & 0.956 & 0.967 & 0.951 & 0.963 & 0.013 & 0.012 \\
\addlinespace[2pt]
\multirow{3}{*}{$\alpha_n=2n^{-2/3}$}
 & 2000 & 0.013 & 0.020 & 0.939 & 0.941 & -- & -- & 0.973 & 0.976 & 0.023 & 0.021 \\
 & 5000 & 0.007 & 0.012 & 0.880 & 0.891 & -- & -- & 0.971 & 0.972 & 0.013 & 0.012 \\
 & 10000 & 0.004 & 0.009 & 0.929 & 0.932 & -- & -- & 0.996 & 0.994 & 0.012 & 0.011 \\
\bottomrule
\end{tabular}

\vspace{2pt}
\parbox{\linewidth}{\footnotesize
\textit{Note.} Density-ratio, Rank-based and Tail report coverage probabilities using the slope estimators $\widehat r_{d,n}$, $\widehat r_{r,n}$ and $\widehat r_{t,n}$, respectively; ``--'' indicates that $\widehat r_{r,n}$ is undefined because $\alpha_n\le b_n$. Length reports the average interval length for the tail estimator $\widehat r_{t,n}$.
}
\end{table}

\paragraph*{Tail inference.}
We evaluate the empirical coverage probabilities of the confidence intervals constructed using the three slope estimators $\widehat r_{d,n}(\alpha_n)$, $\widehat r_{r,n}(\alpha_n)$, and $\widehat r_{t,n}(\alpha_n)$ across $n\in\{2000,5000,10000\}$ with $p_n=0.5$.
Table~\ref{tab:sim-tail} considers three choices of $\alpha_n\in\{n^{-1/3},1.5n^{-1/2},2n^{-2/3}\}$, representing the overlapping, boundary, and extreme-tail regimes summarized in Table~\ref{tab:slope-domains}, respectively, and reports empirical coverage probabilities for all three estimators along with average interval lengths for the tail estimator.
For $\alpha_n=n^{-1/3}$, the rank-based and tail estimators achieve similar valid coverage, consistent with their asymptotic validity in the overlapping regime.
At $\alpha_n=1.5n^{-1/2}$, both estimators continue to perform well, whereas the density-ratio estimator exhibits substantial undercoverage.
In the more extreme regime $\alpha_n=2n^{-2/3}$, the rank-based estimator becomes undefined because $\alpha_n\le b_n$, and the density-ratio estimator exhibits undercoverage, whereas the proposed tail estimator maintains valid conservative coverage throughout.
Covariate adjustment further reduces interval lengths while preserving coverage.

\section{Real data analysis}
\label{sec:real-data}

In this section, we illustrate the proposed methods using data from the randomized experiment of \citet{asanov2023}, which evaluated online learning interventions among more than 45,000 students across 1,151 high schools in Ecuador.
We focus on the Wave-2 experiment, which randomly assigned students to four groups: placebo, encouragement, lottery, and plan-and-team-up, in an approximate $1{:}2{:}2{:}1$ ratio using stratified randomization.
The placebo group received standard platform access, while the other groups received motivational messages, lottery tickets for completed lessons, or prompts to develop a study plan with a household member, respectively.
We focus on the lottery and placebo groups to examine treatment effects across outcome thresholds, extending the comparison beyond mean differences.

We consider two outcomes: a \emph{study-time index} combining minutes on the platform, active days, and completed lessons, and a \emph{knowledge index} combining four subject tests.
As both indices aggregate heterogeneous measures, their numerical scales lack a direct interpretation.
We therefore retain only the pooled student ranks, discarding the original index values.
We approximate the assignment between the lottery and placebo groups as a two-arm completely randomized experiment.
After excluding students with missing outcomes, the final samples comprise 5,897 students for study time ($n_1=3,936$, $n_0=1,961$) and 5,814 for knowledge ($n_1=3,884$, $n_0=1,930$).

\begin{figure}[!t]
\centering
\includegraphics[width=0.8\textwidth]{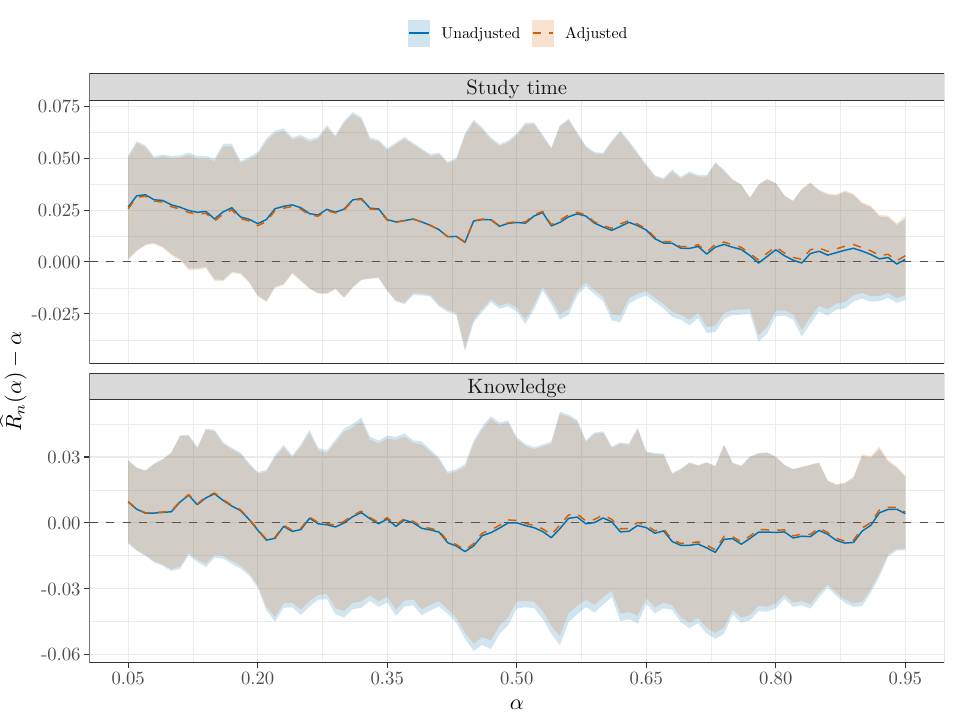}
\caption{
Difference between the estimated ROC curve and the diagonal.
}
\label{fig:real-data-roc}
\end{figure}

We apply the proposed pointwise and simultaneous inference methods with and without covariate adjustment, using baseline platform minutes and active days together with baseline Spanish and English test scores as covariates.
For variance estimation, we use the rank-based slope estimator $\widehat r_{r,n}(\alpha)$ with bandwidth $b_n=n^{-1/2}$, and construct simultaneous confidence bands via Algorithm~\ref{alg:gaussian-band} on the grid $\{0.05,0.06,\ldots,0.95\}$ with $B=50{,}000$ Gaussian draws.
We report deviations from the diagonal, $\widehat\Delta(\alpha)=\widehat R_n(\alpha)-\alpha$, together with 95\% pointwise confidence intervals and simultaneous confidence bands in Table~\ref{tab:real-data} and Figure~\ref{fig:real-data-roc}.
For the study-time index, the lottery produces significant positive deviations at small values of $\alpha$, suggesting that its benefits are more evident among students with higher study-time outcomes, thereby providing insights beyond the mean effect reported by \citet{asanov2023}.
For the knowledge index, both pointwise intervals and simultaneous confidence bands contain zero across all values of $\alpha$ considered.
For both outcomes, covariate adjustment slightly narrows the pointwise intervals and simultaneous confidence bands.

\begin{table}[!htbp]
\centering
\caption{Estimated ROC deviations with pointwise and simultaneous confidence intervals.}
\label{tab:real-data}
\footnotesize
\setlength{\tabcolsep}{3pt}
\begin{threeparttable}
\begin{tabular}{lccccccc}
\toprule
& & \multicolumn{3}{c}{Unadjusted} & \multicolumn{3}{c}{Covariate adjusted} \\
\cmidrule(lr){3-5}\cmidrule(lr){6-8}
Outcome & $\alpha$ & $\widehat\Delta(\alpha)$ & PW 95\% CI & SCB 95\% & $\widehat\Delta_{\mathrm{adj}}(\alpha)$ & PW 95\% CI & SCB 95\% \\
\midrule
\multirow{5}{*}{Study time}
& 0.05 & 0.026 & $[0.010, 0.043]$ & $[0.002, 0.051]$ & 0.026 & $[0.009, 0.042]$ & $[0.001, 0.050]$ \\
& 0.10 & 0.028 & $[0.012, 0.043]$ & $[0.004, 0.051]$ & 0.027 & $[0.011, 0.042]$ & $[0.003, 0.050]$ \\
& 0.25 & 0.026 & $[0.003, 0.049]$ & $[-0.009, 0.061]$ & 0.025 & $[0.003, 0.048]$ & $[-0.009, 0.060]$ \\
& 0.50 & 0.019 & $[-0.009, 0.047]$ & $[-0.024, 0.062]$ & 0.019 & $[-0.008, 0.047]$ & $[-0.022, 0.061]$ \\
& 0.75 & 0.007 & $[-0.014, 0.029]$ & $[-0.026, 0.040]$ & 0.008 & $[-0.012, 0.029]$ & $[-0.023, 0.039]$ \\
\addlinespace
\multirow{5}{*}{Knowledge}
& 0.05 & 0.009 & $[-0.003, 0.022]$ & $[-0.009, 0.028]$ & 0.010 & $[-0.002, 0.022]$ & $[-0.009, 0.028]$ \\
& 0.10 & 0.005 & $[-0.013, 0.023]$ & $[-0.022, 0.032]$ & 0.005 & $[-0.012, 0.022]$ & $[-0.021, 0.032]$ \\
& 0.25 & $-0.003$ & $[-0.029, 0.022]$ & $[-0.042, 0.036]$ & $-0.003$ & $[-0.026, 0.021]$ & $[-0.040, 0.035]$ \\
& 0.50 & 0.000 & $[-0.026, 0.026]$ & $[-0.039, 0.039]$ & 0.001 & $[-0.023, 0.025]$ & $[-0.036, 0.038]$ \\
& 0.75 & $-0.007$ & $[-0.030, 0.015]$ & $[-0.042, 0.027]$ & $-0.006$ & $[-0.028, 0.015]$ & $[-0.040, 0.027]$ \\
\bottomrule
\end{tabular}
\end{threeparttable}
\end{table}

We further examine the upper tail at $\alpha\in\{0.05,0.03,0.01\}$ using the tail inference procedures, without and with covariate adjustment.
As shown in Table~\ref{tab:real-data-tail}, both the unadjusted and covariate-adjusted confidence intervals for the study-time deviation lie above zero at $\alpha=0.05$ and $0.03$, further suggesting that the lottery benefits students with exceptionally high study-time outcomes.
At $\alpha=0.01$, however, both intervals contain zero, providing insufficient evidence of a positive effect at this more extreme threshold.
For the knowledge index, all intervals contain zero, providing no evidence of a positive effect in the upper tail.

\begin{table}[!htbp]
\centering
\caption{Estimated ROC deviations and tail confidence intervals at small $\alpha$.}
\label{tab:real-data-tail}
\footnotesize
\setlength{\tabcolsep}{4pt}
\begin{threeparttable}
\begin{tabular}{lccccc}
\toprule
& & \multicolumn{2}{c}{Unadjusted} & \multicolumn{2}{c}{Covariate adjusted} \\
\cmidrule(lr){3-4}\cmidrule(lr){5-6}
Outcome & $\alpha$ & $\widehat\Delta(\alpha)$ & Tail 95\% CI & $\widehat\Delta_{\mathrm{adj}}(\alpha)$ & Tail 95\% CI \\
\midrule
\multirow{3}{*}{Study time}
& 0.05 & 0.026 & $[0.011, 0.042]$ & 0.026 & $[0.010, 0.041]$ \\
& 0.03 & 0.019 & $[0.005, 0.033]$ & 0.019 & $[0.005, 0.033]$ \\
& 0.01 & 0.007 & $[0.000, 0.015]$ & 0.007 & $[-0.001, 0.014]$ \\
\addlinespace
\multirow{3}{*}{Knowledge}
& 0.05 & 0.009 & $[-0.003, 0.022]$ & 0.010 & $[-0.003, 0.022]$ \\
& 0.03 & 0.007 & $[-0.006, 0.020]$ & 0.007 & $[-0.006, 0.020]$ \\
& 0.01 & $-0.002$ & $[-0.009, 0.006]$ & $-0.001$ & $[-0.008, 0.006]$ \\
\bottomrule
\end{tabular}
\end{threeparttable}
\end{table}

\bibliography{roc}

\end{document}